\documentclass[preprint,
12pt]{elsarticle}
\usepackage{bm}
\usepackage{amsmath}    
\DeclareMathOperator{\sech}{sech}

\usepackage{amssymb}    
\usepackage{graphicx}   
\usepackage{color}
\usepackage[hang]{subfigure}
\newcommand{\jun}{junction }
\newcommand{\juns}{junctions }
\newcommand{\Jos}{Josephson }
\newcommand{\elli}{elliptic }
\newcommand{\ann}{annular }
\newcommand{\conf}{confocal }

\begin{document}

\begin{frontmatter}
\title{Traveling Electromagnetic Waves in \\ Annular \Jos Tunnel Junctions}
\author{Roberto Monaco}
\address{CNR-ISASI, Institute of Applied Sciences and Intelligent Systems ''E. Caianello'', Comprensorio Olivetti, 80078 Pozzuoli, Italy and\\ International Institute for Advanced Scientific Studies (IIASS), Vietri sul Mare, Italy}
\ead{r.monaco@isasi.cnr.it and roberto.monaco@cnr.it}

\date{\today}

\begin{abstract}
It is well known that long \Jos tunnel \juns (JTJs) act as active transmission lines for the slow-mode propagation of magnetic flux-quanta (in the form of solitary waves) that is at the base of many superconducting circuits. At the same time, they support the propagation of quasi-TEM dispersive waves with which the magnetic flux non-linearly interact. In this work, we study the properties of the electromagnetic resonances, under different conditions of practical interest, in annular JTJs (AJTJs), in which the wavelengths are limited to the length of the circumference divided by an integer. Our analysis is based on perturbed sine-Gordon equations the (1+1)-dimensional space with periodic boundary conditions. We discuss the discrete modes of the travelling EM waves in circular annular JTJs in the presence of an in-plane magnetic field, as well as in the recently introduced confocal annular JTJs (in the absence of magnetic field). In both cases, a variable-separation method leads to quantitatively different Mathieu equations characterized by even and odd spatially periodic solutions with different eigenfrequencies. It implicates that a single mode circulating wave is given by the superposition of two standing waves with the same wavelengths but different frequencies, and so has a periodically inverting direction of propagation. The control parameters of this frequency splitting are the in-plane magnetic field amplitude for the circular AJTJ and the aspect ratio for the confocal AJTJs. In the appropriate limits, the previously known solutions are recovered.



\end{abstract}


\end{frontmatter}
%
\tableofcontents
\newpage

\section{Introduction}

Among the many nonlinear equations of physical interest only a few of them, such as, for example, the Korteweg de Vries and the nonlinear Schrodinger equations, possess exact soliton solutions (usually referred to as kinks), i.e., solitary wave packets with permanent profile. One of the solid state systems where soliton static and dynamic properties have been well established is the planar \Jos tunnel \jun (JTJ), a device consisting of two superconducting electrodes separated by a dielectric layer sufficiently thin to allow tunneling of Cooper pairs; more specifically, in a long JTJ, for which one dimension is larger than a characteristic length called Josephson penetration depth, the topological solitons manifest themselves as Cooper pairs current loops, called Josephson vortices or fluxons, as each of them carries one magnetic flux quantum. The leading variable that describes the dynamics of this system is the gauge-invariant phase difference $\phi$ of the wave functions that describe both superconducting electrodes \cite{Joseph64}. The dynamical properties of scalar field $\phi$ are described by a Lorentz-invariant hyperbolic partial differential equation (PDE), named the \textit{sine-Gordon equation}, subjected to appropriate boundary conditions. The properties and applications of the sine-Gordon model, which describes numerous physical systems \cite{Barone71}, are well studied. Over the years, the main reason why JTJs have provided  an important quantitative benchmark for the existence of topological solitons and for the investigation of their dynamics is that real devices are accurately described by nearly-integrable perturbed sine-Gordon equations. In addition, technological progress has enabled the fabrication of various thin-film superconducting systems, with a wide range of electrical parameters and with almost arbitrary geometries. This permits the studies of the influence of nontrivial configurations on the soliton motion. Since the pioneering work of Fulton and Dynes \cite{FD}, the single soliton, which corresponds to a localized $\pm2\pi$-change of $\phi$ that propagates without changing shape, is the solution most investigated in the context of long and narrow JTJs, because a traveling Josephson vortex, due to its unitary topological charge, correspond to a voltage pulse whose spatio-temporal average can be easily monitored in the laboratory. A major experimental manifestation of single and multi-soliton solutions is the emission of electromagnetic radiation by accelerated fluxons \cite{Valery}. However, other non-dispersive solutions to the sine-Gordon equation exist, known in the physical and mathematical literature as  breather solutions, that correspond to bound states of swinging soliton and antisoliton. These localized states, as a result of their zero topological charge, are rather laborious to excite and/or detect in long JTJs \cite{Gulevich12}; another reason is that the soliton-antisoliton oscillations quickly decay with the dissipation always present in real systems. In addition to the above mentioned special solutions to the sine-Gordon equation, there exists a large family of explicit and non-localized solutions, which are periodic in time and space and take the form of nonlinear travelling waves with maximum $\phi$ amplitude equal to $\pi$. They have a continuous spectrum  and a threshold dispersion relation; they are often referred to as \textit{plasma waves}, but, strictly speaking, this terminology should be used only when the oscillation amplitude is small ($|\phi|<<1$) and the waves are harmonic. In the context of JTJs, the non-localized travelling waves describe the propagation of electromagnetic energy with a net power flow. For finite-length JTJs they take the form of pure nonlinear standing waves that can be viewed as the nonlinear interaction of two traveling waves with equal wavelengths and speeds but in opposite directions \cite{Costabile78}. The nonlinear waves can be conveyed into a JTJ by irradiating one of its extremities. In addition, there are several mechanisms by which a moving soliton can generate, often large amplitude, electromagnetic waves: the intrinsic instability of a fluxon moving at relativistic speed, the motion in spatially modulated, or, more generally, in a periodic potential and, under some conditions, the Cherenkov effect \cite{Mints95} . The interaction between a fluxon and the electromagnetic radiation has been the subject of intensive studies over the past three decades \cite{Golubov87,Abdumalikov05,Mussardo06}. Later on, this analysis has been extended to one-dimensional periodic-lattice solutions made of integer \cite{Takayama93} and fractional vortices \cite{Susanto05}. Presently, great interest is being given to the propagation of plasma waves in coupled \Jos \juns aimed to realize oscillators in the terahertz frequency range based on artificially grown or natural layered superconductors \cite{Savel10,Borodianskyi17,Apostolov18}.

\begin{figure}[tb]
\centering
\includegraphics[width=8cm]{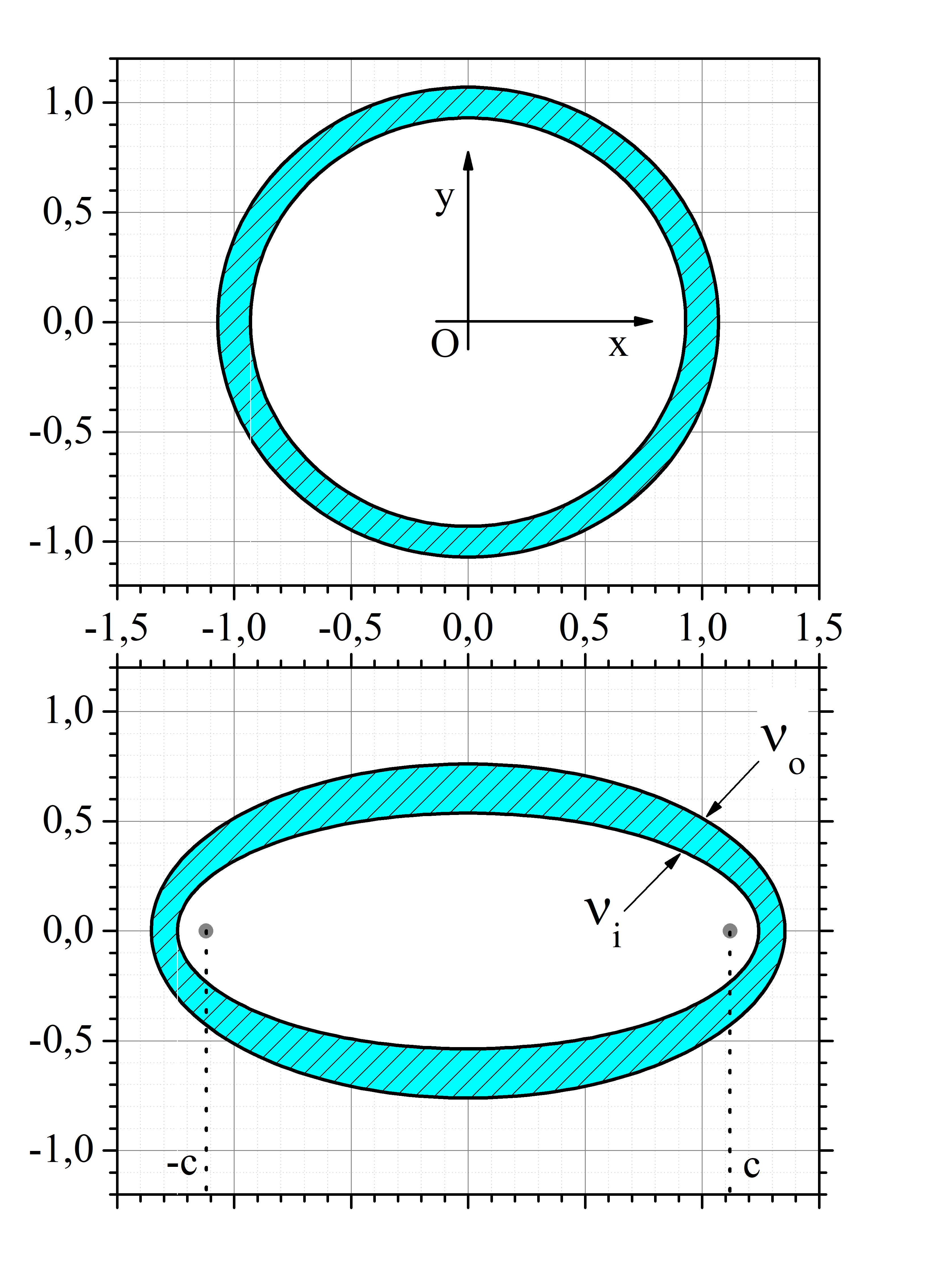}
\caption{(Color online) Top panel: Drawing of the tunneling area of a \textit{circular} annular \Jos tunnel junction (AJTJ) delimited by two closely spaced concentric circles (hatched area); the ring has a unitary mean radius and a constant width. Bottom panel: Drawing of the tunneling area of a \textit{confocal} AJTJ delimited by two closely spaced oblate confocal ellipses (hatched area) identified by the radial elliptic coordinates $\nu_i$ and $\nu_o$. A confocal AJTJ is uniquely characterized by its aspect ratio, $\rho$, and the distance, $2c$, between its foci  (the gray dots); in this specific case $\rho=1/2$ and the foci lie at $(\approx \pm 1.12,0)$ so that the circular and confocal annuli have the same mean circumference (perimeter), $2\pi$. The width, $\Delta w$, of the elliptic annulus varies with the angular elliptic coordinate, $\tau$, as given by Eq.(\ref{width}).}
\label{drawings}
\end{figure}

\medskip

\noindent When studying the propagation of linear and nonlinear waves in JTJs, it is desirable to avoid influences of the boundaries. In this respect, a very interesting object is the \textit{annular} Josephson tunnel junction (AJTJ), topologically obtained by bending a finite length JTJ into a curved form until its extremities are jointed to form a doubly-connected or ring-like JTJ; then the free-ends boundary conditions of the open simply-connected configuration are replaced by periodic conditions. A unique property of AJTJs is the fluxoid quantization in the superconducting loop formed by either the top or the bottom electrodes of the tunnel junction. Then, one or more fluxons may be topologically trapped inside the \jun during the normal-superconducting transition. A \textit{circular} AJTJ, whose tunneling area is sketched in the top panel of Fig.~\ref{drawings}, was first realized in an experiment \cite{Davidson} dating back to 1985 and the existence of trapped fluxon demonstrated. It was soon recognized that AJTJs are unique physical objects for applications of the soliton theory and for the study of the motion of localized excitations without collisions with boundaries, as if it occurred on a infinite transmission line. 

\medskip

\noindent The physics of Josephson tunnel junctions is known to drastically depend on their geometrical configurations and also tiny geometrical details might play a determinant role. Based on the fact that the circle is a special case of an ellipse with no eccentricity, the circular AJTJs have been recently \cite{JLTP16b} generalized to form the so-called \textit{confocal} AJTJ; considering that parallel ellipses do not exist \cite{http1}, the only way to realize an eccentric annulus delimited by an inner and an outer ellipse is when the ellipses have the same foci. The tunneling areas of a confocal AJTJ is drawn by the hatched area in the bottom panel of Fig.~\ref{drawings}. The hatched area is delimited by two closely spaced oblate confocal ellipses whose principal diameters are made parallel to the $X$ and $Y$ axes of a Cartesian coordinate system. The common foci (gray dots) lie on the $X$-axis at $(\pm c,0)$. As the focal points move towards the origin, the eccentricity vanishes and the variable-width confocal annulus progressively reduces to a circular annulus with uniform width. The width of a confocal annulus is smallest at the annulus vertexes (equatorial points) and largest at the co-vertexes (polar points). This smooth variation of the width along the annulus perimeter makes the modeling of a confocal AJTJ very accurate and affords a much richer nonlinear phenomenology \cite{JPCM16}. Both numerical simulations and experiments have proved the existence of solitonic excitations in \conf AJTJS and, more specifically, of a width-induced potential that could be employed for the realization of robust superconducting quantum bits. A strong interaction has also been reported between a travelling fluxon and its self-emitted EM waves although with so far unknown dispersion relation.

\medskip

In this paper we will focus on the properties of the electromagnetic resonances in flux-free AJTJs. The presentation in the paper is organized as follows. Section 2 reviews the propagation of the traveling, in general nonlinear, waves in lossless infinite JTJs. We calculate the energy of the system and study its density dependence. Section 3 addresses the resonant modes in a autonomous circular AJTJ; later on, the analysis is extended to take into account the perturbative effects of  a bias current or an external magnetic field. In Section 4 we present the theoretical model of a confocal AJTJ, based on a perturbed and modified sine-Gordon equation; the wave equation is solved for the small amplitude case and the dispersion relation of the plasma waves is obtained. Section 5 summarizes the results of the analysis and concludes the work.


\section{Periodic travelling waves in a lossless infinite \Jos Transmission Line}

\noindent The quantum-mechanical phase difference, $\phi(\hat{x},\hat{t})$, of a lossless one-dimensional \Jos tunnel \jun (JTJ) obeys a non-linear Klein-Gordon equation \cite{Joseph65,Lebwhol67}, better known as the (1+1)-dimensional \textit{sine-Gordon} equation:

\begin{equation}
\phi_{\hat{x}\hat{x}}- \phi _{\hat{t}\hat{t}} -\sin \phi = 0.
\label{PureSineGordon}
\end{equation}

\noindent Here and in the following, the subscripts on $\phi$ are a shorthand for derivative with respect to the corresponding variable. Furthermore, throughout the paper we use circumflex accents to denote normalized quantities. In Eq.(\ref{PureSineGordon}) $\hat{x}$ is the longitudinal distance normalized to the \textit{\Jos penetration depth}, $\lambda_J=\sqrt{{\Phi_0}/{2\pi \mu_0 J_c d_j}}$, and $\hat{t}$ is the time normalized to the inverse of the so-called (maximum) Josephson \textit{plasma frequency}, $\omega_p\equiv\sqrt{2\pi J_c/\Phi_0 c_s}$ (with $J_c$ being the critical current density, $d_j$ the \textit{current} penetration depth \cite{Wei,SUST13a} and $c_s$ the specific capacitance). It is well known that $\lambda_J$ gives a measure of the distance over which significant spatial variations of the phase occur, in the time independent configuration; the one-dimensional PDE in Eq.(\ref{PureSineGordon})	was derived with the assumption that the transverse dimension, that is, the width of the JTJ is much smaller than $\lambda_J$. In addition, $2\pi/\omega_p$ represents the period of the small-amplitude oscillations in unbiased \textit{small} JTJs \cite{Joseph64} whose dimensions are small compared to $\lambda_J$. Further, we introduce the so called \textit{Swihart velocity}, $\overline{c}\equiv \omega_p\lambda_J$, which, as we will see, gives the speed of light in the barrier \cite{Swihart}.

\noindent On a infinite JTJ, the sine-Gordon equation is completely integrable, with infinitely many conserved quantities; one of its several exact non-trivial solutions to Eq.(\ref{PureSineGordon})  is in the form $\phi(\hat{x}-\hat{v}\hat{t})$ of a periodic waveform travelling with a superluminal propagation velocity $v>\overline{c}$ \cite{Lebwhol67,Barone}: 

\begin{equation}
\phi^{inf}(\hat{x},\hat{t}) = 2\arcsin\!\left\{\kappa\,\text{sn}\!\left[\frac{\hat{x}-\hat{v} \hat{t}}{\sqrt{\hat{v}^2-1}},\kappa^2\right]\right\}= 2\arcsin\!\left\{\kappa\,\text{sn}\!\left[\hat{k}\hat{x}-\hat{\omega}\hat{t},\kappa^2\right]\right\}.
\label{Lebwhol}
\end{equation}

\noindent In Eq.(\ref{Lebwhol}), $\text{sn}\!\left[u,\kappa^2\right]$ is the sine amplitude Jacobian elliptic function of the real variable $u$ and modulus $\kappa^2\leq 1$; furthermore, $\hat{k}\equiv1/\sqrt{\hat{v}^2-1}$ and $\hat{\omega}\equiv\hat{v}/\sqrt{\hat{v}^2-1}$ are two normalized parameters determined by the normalized wave velocity, $\hat{v}\equiv v/\overline{c}$, such that $\hat{\omega}/\hat{k}=\hat{v}$ and $\hat{\omega}^2=1+\hat{k}^2$. As the function $\text{sn}\!\left[u,\kappa^2\right]$ is periodic and $-1\leq\text{sn}\!\left[u,\kappa^2\right]\leq1$, the phase profile in Eq.(\ref{Lebwhol}) describes an infinite series of alternating positive and negative pulses which maintains its shape while translating to the right with phase velocity $\hat{v}\geq1$. This periodic lattice has no internal degrees of freedom and the pulses amplitude is  $A(\kappa)=2\arcsin\, \kappa\leq\pi$. Furthermore, $\text{sn}\!\left[u,\kappa^2\right]$ has a period $4 \text{K}(\kappa^2)\geq2\pi$, where $\text{K}(\kappa^2)\equiv{\int}^{\pi/2}_{0} dz/\sqrt{1-\kappa^2 \sin^2 z}$ is the {\it complete} \elli integral of the first kind. Therefore, the superluminal wave in Eq.(\ref{Lebwhol}) has amplitude-dependent wavelength, $\hat{\lambda}(\kappa)=4\text{K}(\kappa^2)/\hat{k}$, and oscillation period, $\hat{T}(\kappa)=\hat{\lambda}/\hat{v}=4\text{K}(\kappa^2)/\hat{\omega}$, so that $\phi^{inf}(\hat{x}+\hat{\lambda},\hat{t}+\hat{T}) = \phi^{inf}(\hat{x},\hat{t})$. In passing, since $K(0)=\pi/2$, we observe that the oscillation period, $\hat{T}$, depends on the oscillation amplitude, $A$, exactly as for a simple pendulum, namely, $\hat{T}(A)/\hat{T}(0)=(2/\pi)K[\sin^2(A/2)]$. If the wave amplitude, $A$, and wavelength, $\hat{\lambda}$, are fixed, we first determine the corresponding elliptic modulus, $\kappa^2=\sin^2 A/2$, and then the angular wave-number, $\hat{k}=4\text{K}(\kappa^2)/\hat{\lambda}$, the phase velocity $\hat{v}=\sqrt{1+1/\hat{k}^2}$ and the oscillation period. Let us stress that the elliptic modulus, $\kappa$, also parametrizes the pulse density in the train; as $\kappa$ increases in the [0,1] interval, the wave amplitude increases while its density decreases. In the limit of a infinite wavelength ($\hat{k}\to 0$ and $\hat{v} \to \infty$), $\phi^{inf}$ becomes space-independent and the phase uniformly oscillates in time with a frequency $\omega=\omega_p$, no matter what the oscillation amplitude is. Vice-versa, as the wavelength is reduced, the dispersion relation  $\hat{\omega}(\hat{k})$ becomes linear as $\hat{k}>>1$; this implies that in the high frequency limit, the wave becomes non-dispersive and both the phase velocity, $\omega/k$, and the group velocity, $d\omega/dk$ converge to $\overline{c}$, as expected for light waves. Note that Eq.(\ref{Lebwhol}), in which a phase constant was omitted, represents a specific solution to Eq.(\ref{PureSineGordon}) with the initial condition $\phi^{inf}(0,0)=0$, as $\text{sn}\!\left[0,\kappa^2\right]=0$. The stability of the dynamic steady-state solutions in Eq.(\ref{Lebwhol}) and other periodic sine–Gordon traveling waves in a infinite domain was studied in Refs. \cite{Scott69,Marangell13}.

\medskip

\noindent In the (1+1)-dimensional space the Hamiltonian density of a lossless unperturbed sine-Gordon system is \cite{Scott}:

\begin{equation}
\hat{\mathcal{H}}(\hat{x},\hat{t},\phi,\phi_{\hat{x}},\phi_{\hat{t}})=\frac{1}{2}\phi_{\hat{x}}^2 + \frac{1}{2} \phi^2_{\hat{t}}+1-\cos\phi.
\label{hamiltonian}
\end{equation}

\begin{figure}[tb]
\centering
\includegraphics[height=6cm,width=10cm]{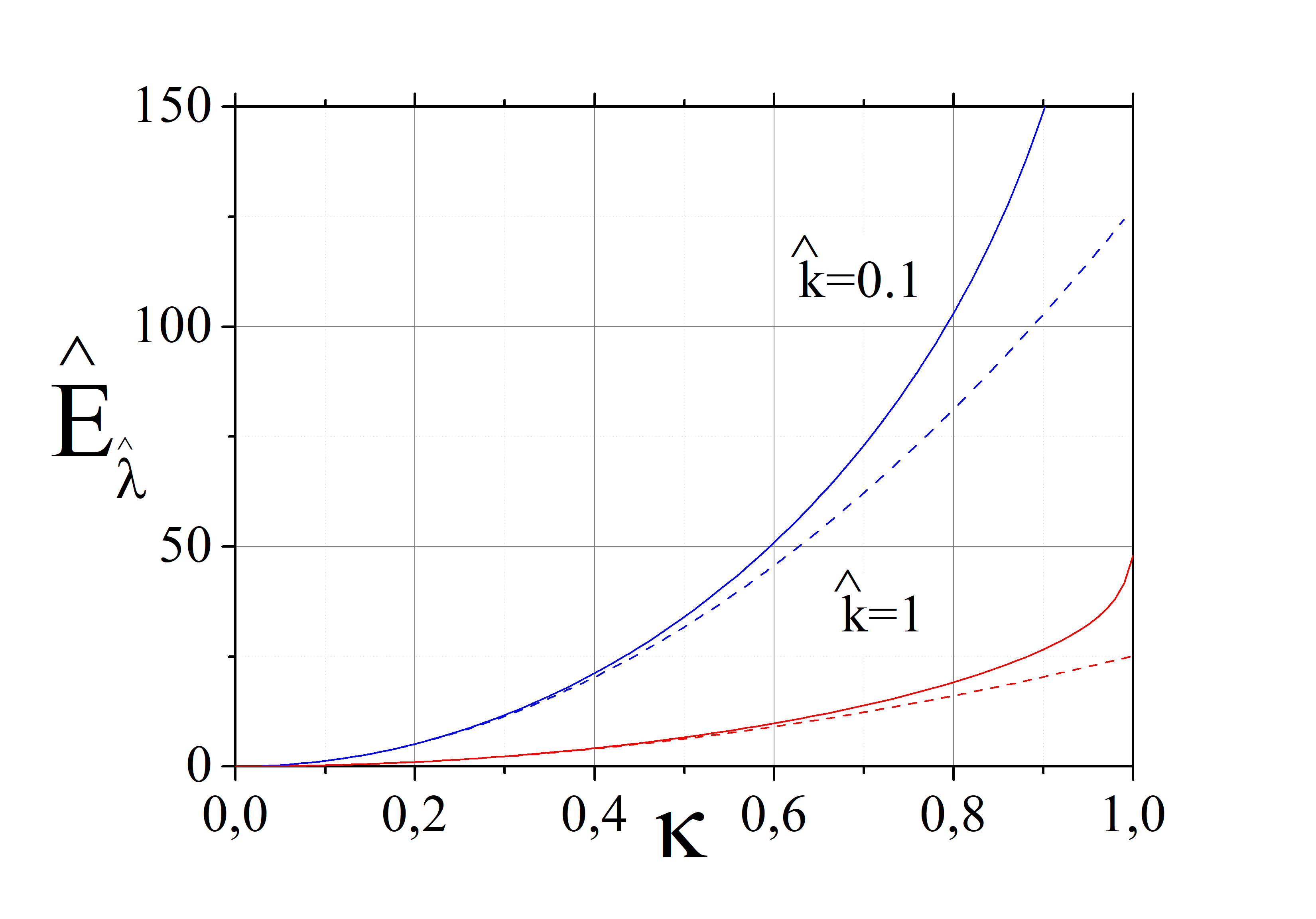}
\caption{(Color online) The wave energy per unit wavelength $\hat{E}_{\hat{\lambda}}$ vs. the elliptic modulus $\kappa$ for two values of the angular wave-number $\hat{k}$; the solid lines show the energies per wavelength according to Eq.(\ref{energyinf}), while the dashed lines follow the approximate quadratic expression for small $\kappa$, i.e., for small amplitude waves, in Eq.(\ref{energyinfapprox}).}
\label{EnergyVsk}
\end{figure}

\noindent Inserting Eq.(\ref{Lebwhol}) in Eq.(\ref{hamiltonian}) and taking into account Eq.(\ref{PureSineGordon}), the Hamiltonian density for our infinite JTJ can be expressed in terms of the cosine amplitude Jacobian elliptic function, $\text{cn}\!\left[u,\kappa^2\right]$, as $\hat{\mathcal{H}}^{inf}(\hat{x},\hat{t})=2\kappa^2 \left\{1+2\hat{k}^2 \text{cn}^2\!\left[\hat{k}\hat{x}-\hat{\omega}\hat{t},\kappa^2\right]\right\}$ whose spatial primitive is:


\begin{equation}
\hat{P}(\hat{x},\hat{t})\equiv\int \hat{\mathcal{H}}^{inf}(\hat{x},\hat{t})\, d\hat{x}=2 \kappa^2 \hat{x} + 4 \hat{k} (\kappa^2-1) (\hat{k} \hat{x}-\hat{\omega} \hat{t})+ 4 \hat{k} \text{E}[\text{Am}[\hat{k} \hat{x}-\hat{\omega} \hat{t}, \kappa^2], \kappa^2],
\label{primitive}
\end{equation}

\noindent where $\text{E}[u,\kappa^2]\equiv{\int}^{u}_{0} \sqrt{1-\kappa^2 \sin^2 z} dz$ is the \elli integral of the second kind and $\text{Am}[u,\kappa^2]$ is the inverse function of the elliptic integral of the first kind also called the Jacobi amplitude. The energy stored per wavelength $\hat{\lambda}$ is: 

\begin{equation}
\hat{E}_{\hat{\lambda}}(\kappa,\hat{k})\equiv\int^{\hat{x}_0+\hat{\lambda}}_{\hat{x}_0}\hat{\mathcal{H}}^{inf} d\hat{x}=\hat{P}[\hat{x}_0+4\text{K}(\kappa^2)/\hat{k},\hat{t}]-\hat{P}(\hat{x}_0,\hat{t}),
\label{energy}
\end{equation}

\noindent with arbitrary $\hat{x}_0$. Inserting Eq.(\ref{primitive}) in Eq.(\ref{energy}) and using the identities $\text{Am}[a+4\text{K}(\kappa^2), \kappa^2]=\text{Am}[a,\kappa^2]+2\pi$ and $\text{E}[a+2\pi,\kappa^2]=\text{E}[a,\kappa^2]+4\text{E}(\kappa^2)$, where $\text{E}(\kappa^2) \equiv \text{E}[2\pi,\kappa^2]$ is the {\it complete} \elli integral of the second kind of argument $\kappa^2$, we end up with the energy per unit wavelength:

\begin{equation}
\hat{E}_{\hat{\lambda}}(\kappa,\hat{k})=\frac{8 \kappa^2}{\hat{k}}(1+2\hat{k}^2)\text{K}(\kappa^2)-16\hat{k}[\text{K}(\kappa^2)-\text{E}(\kappa^2)],
\label{energyinf}
\end{equation}

\noindent that is an integral of motion ($d\hat{E}_{\hat{\lambda}}/d{\hat{t}}=0$). Then, the wave amplitude, through the squared modulus, $\kappa^2$, determines the wave energy density. Seen from the opposite point of view, the energy density of the superluminal wave (together with the free parameter $\hat{v}$) parametrizes different solutions to Eq.(\ref{Lebwhol}). The solid lines in Fig.~\ref{EnergyVsk} show the monotonic increase of $\hat{E}_{\hat{\lambda}}$ with $\kappa$ for two values of the wave-number $\hat{k}$ according to Eq.(\ref{energyinf}).

\subsection{Small amplitude waves}

\noindent In the limit of small amplitude waves, $\kappa <<1$, being $\text{sn}\!\left[u,\kappa^2\right]\approx\sin u$, $A/2=\arcsin\,\kappa\approx \kappa$ and $K(\kappa^2)\approx\pi/2$, the permanent-shape of the progressive wave in Eq.(\ref{Lebwhol}) reduces to the one-dimensional plane wave:

\begin{equation}
\phi^{inf}_0(\hat{x},\hat{t}) = A \sin(\hat{k}_0\hat{x}-\hat{\omega}_0\hat{t});
\label{planewave}
\end{equation}

\noindent under this circumstance, the wave amplitude, $A$, the wavelength $\hat{\lambda}_0\equiv\hat{\lambda}(0)=2\pi/\hat{k}_0$ and the period $\hat{T}_0\equiv\hat{T}(0)=2\pi/\hat{\omega}_0$ are no longer $\kappa$-dependent. Eq.(\ref{planewave}) is a solution to the Klein-Gordon equation, $\phi_{\hat{x}\hat{x}}- \phi _{\hat{t}\hat{t}}-\phi = 0$; in fact, being $|\phi|<<1$, then $\sin\phi \approx \phi$ in Eq.(\ref{PureSineGordon}). The amplitude-independent dispersion relation, $\hat{\omega}_0^2=1+\hat{k}_0^2$, characterized by a unitary cut-off frequency, is typical of a light wave in a electron plasma \cite{Joseph64}. In different words, $\omega_p$ represents the lowest frequency that allows the propagation of electromagnetic waves inside a JTL.

\medskip

\noindent $\text{E}(0)=\text{K}(0)=\pi/2$, so, if we evaluate the energy per wavelength in Eq.(\ref{energyinf}) for $\kappa=0$, we obtain zero as expected for a zero-amplitude wave. However, to the second order of approximation, for small $\kappa$ it is $\text{K}(\kappa^2)\approx \pi(1+\kappa^2/4)/2$ and $\text{E}(\kappa^2)\approx \pi(1-\kappa^2/4)/2$. Therefore, in this limit, the energy per wavelength reduces to:
\medskip
\begin{equation}
\hat{E}_{\hat{\lambda}}(\kappa,\hat{k}_0) \approx \frac{\pi \kappa^2}{\hat{k}_0}\left[4(1+\hat{k}_0^2)+\kappa^2 (\hat{\omega}^2+\hat{k}_0^2)\right] \approx \frac{4\pi \kappa^2 (1+\hat{k}_0^2)}{\hat{k}_0} 
\label{energyinfapprox}
\end{equation}

\noindent which is an energy that grows quadratically with the wave amplitude, as $4\kappa^2\approx A^2$. The dashed lines in Fig.~\ref{EnergyVsk} follow the approximate quadratic expression for small $\kappa$ in Eq.(\ref{energyinfapprox}); we note that the approximation is accurate for $\kappa\lesssim 0.4$ corresponding to a wave amplitude $A\lesssim 0.8$. The Eq.(\ref{energyinfapprox}) can be cast in the more compact form $\hat{E}_{\hat{\lambda}}(\kappa,\hat{k}_0) \approx \pi A^2 \hat{\omega}_0^2/\hat{k}_0$.

\section{Rotating waves along a circular AJTJ}

The PDE of a circular AJTJ with mean radius $r$ (see the top panel of Figure~\ref{drawings}) in the presence of a magnetic field, $H$, applied in the junction plane is \cite{PRB97}:
\vskip -10pt

\begin{equation}
\left(\frac{\lambda_J}{r}\right)^2  \phi_{\theta\theta} - \phi _{\hat{t}\hat{t}}-\sin \phi = \alpha \phi_{\hat{t}} -\gamma + h\sin(\theta-\bar{\theta}),
\label{PDEring}
\end{equation}


\noindent where $\theta\equiv x/r$ is the angular coordinate and, as in Eq.(\ref{PureSineGordon}),  time is normalized to the inverse of the \jun plasma frequency, $\omega_p$. In the right-hand side of Eq.(\ref{PDEring}) we have grouped the usual perturbative terms \cite{Scott}: the normalized losses due to tunneling of normal electrons across the dielectric barrier, $\alpha \phi_{\hat{t}}$, the normalized uniform bias current density, $\gamma<1$, forced through the junction and the normalized magnetic field, $h\equiv H/J_c r$, applied with and angle $\bar{\theta}$ with respect to the $y$-axis. 

\medskip

\noindent When cooling an annular JTL below its critical temperature one or more flux quanta may be spontaneously trapped in its doubly connected electrodes. The trapping probability is known to increase with the speed of the normal-to-superconducting transition \cite{PRB06,PRB08}. The algebraic sum of the flux quanta trapped in each electrode is an integer number $n_w$, called the winding number, counting the number of Josephson vortices (fluxons) trapped in the \jun barrier. From a mathematical point of view a fluxon is a $2\pi$-kink in the \Jos phase within a distance of approximately one \Jos penetration depth. To take into account the number of trapped fluxons, Eq.(\ref{PDEring}) is supplemented by the periodic boundary conditions \cite{PRB97}:

\begin{subequations}
\begin{eqnarray}\label{periodic1}
\phi(\theta+2\pi,\hat{t})=\phi(\theta,\hat{t})+ 2\pi n_w,\\
\phi_\theta(\theta+2\pi,\hat{t})=\phi_\theta(\theta,\hat{t}).
\label{periodic2}
\end{eqnarray}
\end{subequations}

\noindent Throughout this paper we will limit our analysis to the simplest case of no trapped flux, i.e., we will set $n_w=0$. Apart from the viscous dissipative term, $\alpha \phi_{\hat{t}}$, Eq.(\ref{PDEring}) may be derived from the Hamiltonian density:

\begin{equation}
\hat{\mathcal{H}}(\theta,\hat{t},\phi,\phi_{\theta},\phi_{\hat{t}})=\frac{1}{2}\left(\frac{\phi_{\theta}}{\hat{r}}+h\hat{r}\cos\theta\right)^2 + \frac{1}{2}\phi^2_{\hat{t}}+1-\cos\phi+\gamma\phi.
\label{hamiltonianFull}
\end{equation}

\noindent Making use of Eq.(\ref{PDEring}) and of the periodic boundary conditions, it can be shown that the time derivative of the energy is finite and non-positive \cite{Lebwhol67}:

\begin{equation}
\frac{dE}{dt}=\oint \frac{d\hat{\mathcal{H}}}{dt} d\theta=-\alpha \oint \phi^2_{\hat{t}} d\theta.
\label{EnerDer}
\end{equation}

\noindent The rate of dissipation of energy is due to the shunt loss, i.e., Ohmic currents through  the barrier. Clearly, the losses do not affect the static solutions; their effect on a time-oscillating solution is to exponentially reduce its oscillation amplitude. In this paper, in order to carry out analytical derivations, we assume an underdamped regime that reduces Eq.(\ref{PDEring}) to:

\begin{equation}
\frac{1}{\hat{r}^2}\phi_{\theta\theta} - \phi _{\hat{t}\hat{t}}-\sin \phi \approx -\gamma + h\sin(\theta-\bar{\theta}),
\label{PDEring1}
\end{equation}

\noindent where we have introduced the normalized radius $\hat{r}\equiv r/\lambda_j$. The implications of the periodic conditions on the travelling waves allowed in a circular AJTJ are first examined in the absence of external bias current and magnetic field. Then Eq.(\ref{PDEring1}) becomes:

\begin{equation}
\frac{1}{\hat{r}^2}  \phi_{\theta\theta}- \phi _{\hat{t}\hat{t}} -\sin \phi = 0.
\label{SGring}
\end{equation}

\noindent that is identical to  Eq.(\ref{PureSineGordon}) and therefore has similar solutions. In order to find the solutions to the autonomous PDE in Eq.(\ref{SGring}) that fulfill the boundary conditions in Eqs.(\ref{periodic1}) and (\ref{periodic2}) with $n_w=0$, we have to select those progressive waves in Eq.(\ref{Lebwhol}) whose wavelengths, $\lambda_m$, are sub-multiples of the ring circumference, $\lambda_m\equiv\hat{\lambda}_m \lambda_J=2\pi r/m$, with $m$ being a positive integer representing the number of oscillations (nodes) accommodated along the circle. The wave-numbers and frequencies associated with the mode $m$ are, respectively, $\hat{k}_m(\kappa)=4 K(\kappa^2)/\hat{\lambda}_m=4 m  K(\kappa^2)/(2\pi\hat{r})$ and $\hat{\omega}_m(\kappa)=\sqrt{1+\hat{k}^2_m(\kappa)}$. Recalling that $\hat{x}=\hat{r}\theta$, the discrete solutions to Eq.(\ref{SGring}) are:

\begin{equation}
\phi^{circ}_m(\theta,\hat{t}) = 2\arcsin\!\left\{\kappa\,\text{sn}\!\left[\hat{k}_m\hat{x}(\theta)-\hat{\omega}_m\hat{t},\kappa^2\right]\right\}= 2\arcsin\!\left\{\kappa\,\text{sn}\!\left[\hat{k}_m \hat{r}\theta-\hat{\omega}_m\hat{t},\kappa^2\right]\right\}.
\label{circular}
\end{equation}

\noindent Eq.(\ref{circular}) represents constant-shape periodic waves rotating clockwise with phase velocity $\hat{v}_m(\kappa)=\hat{\omega}_m(\kappa)/\hat{k}_m(\kappa)$ and oscillating about $\phi=0$ with period $\hat{T}_m=4 K(\kappa^2)/\hat{\omega}_m$. As before, the wave amplitude, $A$, of a rotating wave train determines the elliptic modulus, $\kappa=\sin A/2$. It is worth to stress that, as the product $\hat{k}_m \hat{r}=2 m  K(\kappa^2)/\pi$ is independent on $\lambda_J$, our  analysis is not restricted to electrically long AJTJs. In the small amplitude limit, the $m$-mode of the uniform rotational motion in Eq.(\ref{circular}) reduces to a rotating linear wave:

\begin{equation}
\phi^{circ}_m(\theta,\hat{t}) = A_m \sin\left(m\theta-\hat{\omega}_m\hat{t}\right),
\label{circularsmall}
\end{equation}

\noindent where $\hat{\omega}_m^2=1+ (m/\hat{r})^2$ and $|A_m|<<1$. The major difference between the solutions in Eq.(\ref{Lebwhol}) and Eq.(\ref{circular}) resides in their spectra, which are continuous for the former and discrete for the latter. The same reasoning applies to their small-amplitude counterparts, respectively, in Eq.(\ref{planewave}) and Eq.(\ref{circularsmall}).



\subsection{Small amplitude oscillations around a static solution}

Let $\psi(\theta)$ be a static solution to the more general Eq.(\ref{PDEring1}). The task of finding the small amplitude electromagnetic waves $\tilde{\phi}(\theta,\hat{t})$ that oscillate around $\psi(\theta)$ can be achieved adopting a perturbative approach in which the right-hand side of Eq.(\ref{PDEring1}) is considered as a disturbance. If we assume that the solutions to a perturbed system are close to the corresponding solutions to the unperturbed (integrable) system, then one calculates the deviation of the perturbed solution from the unperturbed one . Accordingly, we may insert the ansatz $\phi(\theta,\hat{t}) \approx \psi(\theta)+\tilde{\phi}(\theta,\hat{t})$ into Eq.(\ref{SGring}) yielding \cite{Dahm68,Solymar}:

\begin{equation}
\frac{1}{\hat r^2}  \tilde{\phi}_{\theta\theta}- \tilde{\phi} _{\hat{t}\hat{t}} - \cos \psi(\theta) \,\tilde{\phi}=0,
\label{perturbRing}
\end{equation}

\noindent where we considered $|\tilde{\phi}(\theta,\hat{t})|<<1$ to be a small perturbation of the reigning static condition. We look for clockwise rotating non-dissipative waves of the form $\tilde{\phi}(\theta,\hat{t})=  \tilde{\Phi}(\theta) \exp[-\hat{\imath} (\hat{\omega} \hat{t}-\delta)]$, where the time dependence has been expressed in complex form. Inserting $\tilde{\phi}$ into Eq.(\ref{perturbRing}), we get a linear second-order homogeneous ordinary differential equation (ODE) for the (in general, complex) function  $\tilde{\Phi}(\theta)$:

\vskip -8pt
\begin{equation}
-\frac{1}{\hat r^2}  \tilde{\Phi}_{\theta\theta} + \cos \psi(\theta) \,\tilde{\Phi}= \hat{\omega}^2 \tilde{\Phi}.
\label{schr}
\end{equation}

\noindent Without loss of generality, the (real) phase constant, $\delta$, will be chosen such that $Re[\tilde{\phi}(0,0)]=0$, to be consistent with the previously found solutions. Eq.(\ref{schr}) resembles the time-independent one-dimensional Schr\"{o}dinger-equation for an electron with complex wave-function $\tilde{\Phi}$ and positive energy $\hat{\omega}^2$ moving in a periodic potential $\cos \psi(\theta)$. The boundary condition to be imposed on $\tilde{\Phi}$ is that it be single valued in $\theta$, that is, $\tilde{\Phi}(\theta+2\pi)=\tilde{\Phi}(\theta)$. In the remaining part of this Section we will seek the periodic solutions to Eq.(\ref{schr}) in two specific static cases where either a bias current or a magnetic field is applied. This will allow, in turn, to go back to the solutions to Eq.(\ref{PDEring1}) that oscillate around the static solution $\psi(\theta)$.

\subsubsection{$\gamma\neq0$ and $h=0$}

\noindent The simplest case occurs when a homogeneous bias current, $\gamma$, is applied to the circular AJTJ in the absence of a magnetic field. Then, $\psi_\gamma=\arcsin\gamma$ is the (uniform) static solution to the Eq.(\ref{PDEring1}) with $h=0$. Since $\cos \psi_\gamma= \sqrt{1-\gamma^2}$ is spatially independent, then Eq.(\ref{schr}) is a linear ODE with constant coefficients whose complex periodic solutions are in the form $\tilde{\Phi}_{m,\gamma}(\theta)=A_m \exp \hat{\imath} m\theta$ with integer $m$ and $|A_m|<<1$, provided that $m^2/\hat{r}^2=\hat{\omega}_m^2-\sqrt{1-\gamma^2}$. By taking the real part of $\tilde{\Phi}_{m,\gamma}(\theta) \exp[-\hat{\imath} (\hat{\omega} \hat{t}-\delta)]$, the solutions to Eq.(\ref{perturbRing}) satisfying the periodic conditions in Eqs.(\ref{periodic1}) and (\ref{periodic2}) (with $n_w=0$) are in the form of $m$-th order rotating plane waves:

\vskip -8pt
\begin{equation}
\tilde{\phi}_{m,\gamma}(\theta,\hat{t})= A_m \sin(m\theta-\hat{\omega}_{m,\gamma}\hat{t}),
\label{planewavecirc}
\end{equation}

\noindent where $\hat{\omega}_{m,\gamma}^2= \hat{k}_m^2+ \sqrt{1-\gamma^2}$ with $\hat{k}_m\equiv 2\pi/\hat{\lambda}_m=m/\hat{r}$; this implies that the wave frequencies, as well as the phase velocities, $\hat{\omega}_{m,\gamma}/\hat{k}_m$, decrease with the bias amplitude. In Eq.(\ref{planewavecirc}) the phase constant, $\delta$, was set equal to $\pi/2$, in order to have $\tilde{\phi}_{m,\gamma}(0,0)=0$. The harmonic field in Eq.(\ref{planewavecirc}) describes small oscillations about the static solution $\psi_\gamma=\arcsin\gamma$ and reduces to Eq.(\ref{circularsmall}) for $\gamma=0$.  For $m=0$, we have a spatially-independent oscillation with frequency $\omega_{0,\gamma}=\omega_p \sqrt[4]{1-\gamma^2}$, as already found for a biased point-like JTJ \cite{Ivanchenko,Ruggiero95}. In passing, one observes that, by inserting $\phi_{m,\gamma}= \psi_\gamma + \tilde{\phi}_{m,\gamma}$ into the energy density of Eq.(\ref{hamiltonian}), it ends up with a constant energy $E_{m,\gamma}\equiv\oint \hat{\mathcal{H}}d\theta=2\pi \hat{r}(A_m^2 \hat{\omega}_m^2+1-\sqrt{1-\gamma^2})/m$. This means that a bias current does not supply any power to the plasma waves, at variance with what occurs to the solitary waves.
\medskip



%
%

\subsubsection{$\gamma=0$ and $h\neq0$}
\label{sec:sexion}
\medskip
\noindent We next consider the more interesting case of an unbiased circular AJTJ in the presence of a small normalized in-plane magnetic field, $|h|<<1$, applied in the $\bar{\theta}$ direction. On neglecting terms of order of $h^2$ and above, the static solution to Eq.(\ref{PDEring1}) (with $\gamma=0$) is \cite{notePsih} $\psi_h(\theta) = 2\arcsin\,\left[(h_e/2) \sin(\bar{\theta}-\theta)\right]$ with an effective magnetic field, $h_{e}\equiv h \hat{r}_e^2$, proportional to the square of an effective normalized radius $\hat{r}_e\equiv\hat{r}/\sqrt{1+\hat{r}^2}$ (note that $\hat{r}_e\approx 1$ and $h_{e}\approx h$ for rings with very long normalized perimeter). Due to the symmetry of the system we can assume, without any loss of generality, that the  magnetic field is applied along the $y$-axis, so that $\bar{\theta}=0$ in Eq.(\ref{PDEring}).

\noindent The search for the wave-like solutions, $\tilde{\phi}_{m,h}(\theta,\hat{t})$, of Eq.(\ref{perturbRing}) requires some additional considerations. Being $\cos \psi_h(\theta)= 1-(h_e/2)^2+(h_e/2)^2 \cos 2 \theta$, then Eq.(\ref{schr}) for $\tilde{\Phi}_h(\theta)$ can be rearranged as: 

%
\begin{equation}
 \frac{d^2\tilde{\Phi}_h}{d\theta^2} + \left(a-2q\cos2\theta\right) \tilde{\Phi}_h(\theta) =0,
\label{mathieu0}
\end{equation}

\noindent where $a=\hat{r}^2(\hat{\omega}^2+h_e^2/4-1)$ and $q= \hat{r}^2 h_e^2/8\geq0$. Eq.(\ref{mathieu0}) is a linear ODE named the \textit{angular Mathieu differential equation} whose general solutions (for given $a$ and $q$) are linear combinations of a even function, $C(a,q,\theta)$, called \textit{Mathieu cosine} and a odd function, $S(a,q,\theta)$, called \textit{Mathieu sine}, namely, $\tilde{\Phi}_h(a,q,\theta)=A C(a,q,\theta)+B S(a,q,\theta)$, where $A$ and $B$ are constants depending on the initial conditions. We observe that the parameters $a$ and $q$ merely depend on the magnetic field amplitude and not on its sign. For $q=0$ (that is for $h=0$), as expected, the Mathieu cosine and sine functions reduce to $\cos \sqrt{a_0} \theta$ and $\sin \sqrt{a_0} \theta$, respectively, with $\sqrt{a_0}=\pm \hat{r}\sqrt{\hat{\omega}^2-1}=\pm m$. For nonzero $q$, the Mathieu functions have a fairly complicated behavior and are periodic for only certain values of $a$. Given $q$, for countably many special values of $a$, called \textit{characteristic values} or \textit{eigenvalues}, the Mathieu equation admits solutions which are periodic with angular period $2\pi/m$, where $m=1,2,3,..$ is a positive integer. The characteristic values of the Mathieu cosine and sine functions are written, respectively, $a_m(q)$ and $b_m(q)$. They are given by the Wolfram Language functions \textsl{MathieuCharacteristicA}$[r, q]$ and \textsl{MathieuCharacteristicB}$[r,q]$ with $r$ an integer or a rational number. For positive $q$, it is $a_1 < b_1 <b_2 <a_2 <a_3 <b_3 < b_4 ....$. Since the characteristic numbers $a_m$ and $b_m$ are not equal, $C(a_m,q,\theta)$ and $S(b_m,q,\theta)$ are not two independent solutions to the same Mathieu equation. In fact, the second solution corresponding to $C(a_m,q,\theta)$, namely $S(a_m,q,\theta)$, is a nonperiodic function, as is also the second solution, $C(b_m,q,\theta)$, corresponding to $S(b_m,q,\theta)$. In passing, we note that $C(a_m,q,\theta)$ and $S(b_m,q,\theta)$ are real functions with the same periodicity, but opposite parity; therefore, they are orthogonal. As the non-periodic solutions are excluded by our periodicity conditions, for a given $q$ (i.e., for a given magnetic field value $h$), as a first step, we can choose between either a periodic even solution, $\propto C(a_m,q,\theta)$, or an odd one, $\propto S(b_m,q,\theta)$. 


\medskip

\noindent First, let us select the $m$-th even solution $\tilde{\Phi}_{h,m}^e=A_{m} C(a_m,q,\theta)$; then, in order for $\tilde{\Phi}_{h,m}^e(\theta) \exp[-\hat{\imath} (\hat{\omega} \hat{t}-\delta)]$ to be a solution to Eq.(\ref{schr}), it must be $\hat{\omega}(a_m)\equiv\hat{\omega}_{a,m} = \sqrt{1+a_m/\hat{r}^2-(h_e/2)^2}$. The real part of this solution, $\tilde{\phi}^e_{m,h}(\theta,\hat{t})=A_m C(a_m,q,\theta) \sin\hat{\omega}_{a,m}\hat{t}$, represents a standing, rather than a traveling, wave (the phase constant $\delta$ was set to $\pi/2$, as we wanted $\tilde{\phi}^e_{m,h}(0,0)=0$). Vice-versa, in case we opt for the $m$-th odd solution, $\tilde{\Phi}_{h,m}^o=B_{m} S(b_m,q,\theta)$ with $|B_m|<<|h|$, we have different frequencies $\hat{\omega}_{b,m} \equiv \sqrt{1+b_m/\hat{r}^2-(h_e/2)^2}$; this latter solution results in a second standing wave initially shifted by $\pi/2$ with respect to the former one, namely, $\tilde{\phi}^o_{m,h}(\theta,\hat{t})=B_m S(b_m,q,\theta) \cos\hat{\omega}_{b,m}\hat{t}$. As $\tilde{\phi}^o_{m,h}$ and $\tilde{\phi}^e_{m,h}$ are independent solutions with the same angular periodicity $2\pi/m$ of the linear PDE in Eq.(\ref{perturbRing}) in the presence of a small normalized magnetic field, $h$, the general solution, $\tilde{\phi}_{m,h}$, with period $2\pi/m$ is the superposition:

\vskip -8pt
\begin{equation}
\tilde{\phi}_{m,h}(\theta,\!\hat{t})\!\!=\!\!\tilde{\phi}^o_{m,h}(\theta,\!\hat{t})+\tilde{\phi}^e_{m,h}(\theta,\!\hat{t})\!\!=\!\!A_m C(a_m,\!q,\!\theta)\!\sin\hat{\omega}_{a,m}\hat{t}+B_m S(b_m,\!q,\!\theta)\!\cos\hat{\omega}_{b,m}\hat{t}.
\label{phimh}
\end{equation}

\begin{figure}[tb]
\centering
\includegraphics[height=6cm,width=10cm]{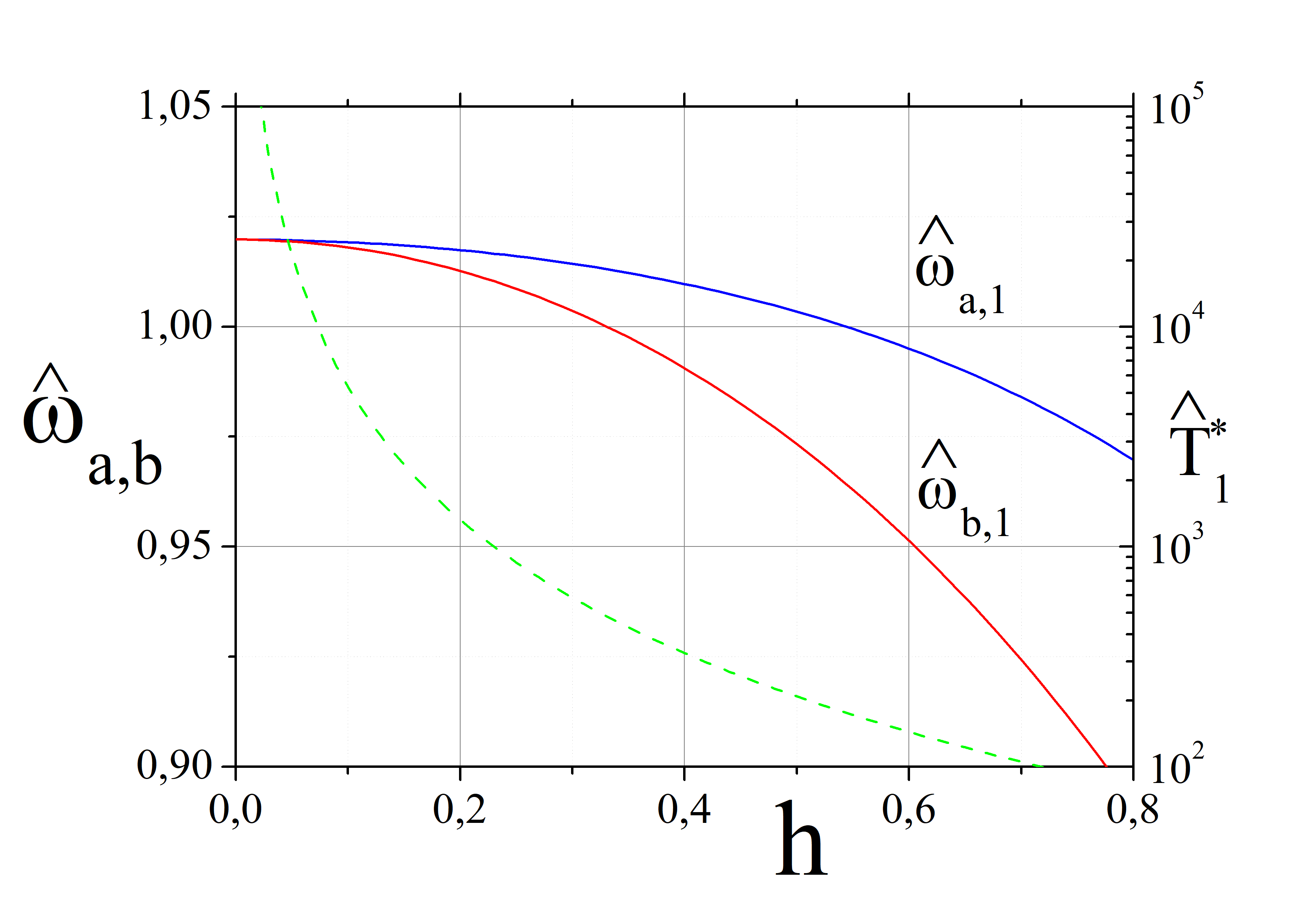}
\caption{(Color online) The solid lines show the magnetic field dependence of the first-mode eigenfrequencies, $\hat{\omega}_{a,1}$ and $\hat{\omega}_{b,1}$, for a circular AJTJ with a normalized radius $\hat{r}=5$; for $h=0$, $\hat{\omega}_{a,1}=\hat{\omega}_{b,1}=\hat{\omega}_1\equiv \sqrt{1+(1/\hat{r})^2}=1.0198$. The dashed line represents the  wave periodicity, $\hat{T}_1^*(h)$, and refers to the right (logarithmic) scale.}
\label{omegaab}
\end{figure}

\noindent Here, $\tilde{\phi}_{m,h}(\theta,\hat{t})$ is a linear combination of two standing waves with incommensurate frequencies ($\hat{\omega}_{a,m} > \hat{\omega}_{b,m}$ for $h\neq0$); therefore, its time evolution is rather intricate, because the resulting wave not only does not maintain its shape, but also periodically changes its direction of propagation. At $\hat{t}=0$ the phase difference of the two standing waves equals $\pm \pi/2$ depending on the relative sign between the amplitudes $A_m$ and $B_m$ and their sum results in a wave rotating clockwise (counterclockwise), if the amplitudes have opposite (same) sign. The original relative phase will be restored after a period of time, $\hat{T}_m^*$, such that $\hat{\omega}_{a,m}\hat{T}_m^*=\hat{\omega}_{b,m}\hat{T}_m^*+2\pi$, that is $\hat{T}_m^*=2\pi/(\hat{\omega}_{a,m}-\hat{\omega}_{b,m})$. In different words, the phase difference changes by $\pi/2$ every $\hat{T}_m^*/4$. Then, for $\hat{t}$ near $\hat{T}_m^*/4$ and $3\hat{T}_m^*/4$ the two standing waves are either in phase or out of phase and their sum results in another standing wave; for $\hat{t}$ near $\hat{T}_m^*/2$ the original phase displacement changes its sign and the wave starts to rotate in the opposite direction. As the characteristic value $a_m$ for the even part of the solutions $\Phi_m(\tau)$ with periodicity $2\pi/m$ depends on the magnetic field through the coefficient $q$, the field dependence of the eigenfrequencies $\hat{\omega}_{a,m}$ where derived by (numerically) finding the root of the equation \textsl{MathieuCharacteristicA}$[m, q(h)]=\hat{r}^2(\hat{\omega}_{a,m}^2-1)+2q(h)$ with $q(h)\equiv h^2 \hat{r}_e^2 \hat{r}^2 /8$. A similar process was performed to determine the eigenfrequencies $\hat{\omega}_{b,m}$ for the odd part of $\Phi_m(\tau)$. (In passing, we note that, for small $q$, the eigenfrequencies can be more easily found by resorting to the series expansions of the Mathieu characteristic numbers \cite{http2,http3}.) Once the eigenfrequencies, $\hat{\omega}_{a,m}$ and $\hat{\omega}_{b,m}$, are found, the eigenvalues, $a_m$ and $b_m$, are derived and the $m$-mode solution with the condition $\tilde{\phi}_{m,h}(0,0)=0$ in Eq.(\ref{phimh}) is fully determined. The solid lines in Fig.~\ref{omegaab} show the field dependence of the frequencies $\hat{\omega}_{a,1}$ and $\hat{\omega}_{b,1}$ for a circular AJTJ with a normalized perimeter $\ell\equiv2\pi\hat{r}=10\pi$, that is, with $\hat{r}=5$. We notice that, for $h=q=0$, being $a_m=b_m=m^2$, the two frequencies become degenerate, $\hat{\omega}_{a,m}=\hat{\omega}_{b,m}=\hat{\omega}_m\equiv \sqrt{1+(m/\hat{r})^2}$. As the magnetic field increases, the eigenfrequencies split and both decrease; at the same time, the wave time periodicity $\hat{T}_1^*(h)$ get smaller, as indicated by the dashed line referred to the right (logarithmic) scale of the plot. Summarizing, the complete phase profile of a circular AJTJ in the presence of a small in-plane magnetic field is: 
$$\phi_{m,h}(\theta,\hat{t}) \approx \psi_h(\theta)+\tilde{\phi}_{m,h}(\theta,\hat{t})=$$
$$= 2\arcsin\,\left[(h_e/2)\sin\theta\right]+A_m C(a_m,q,\theta)\sin\omega_a \hat{t}+B_m S(b_m,q,\theta)\cos\omega_b \hat{t}.$$

\noindent We notice that, with $B_m=-A_m$ and $h=0$, as expected, this expression reduces to the plane wave solution in Eq.(\ref{circularsmall}), namely, $A_m \sin(m\theta-\hat{\omega}_m\hat{t})$. A Mathematica-based program that displays the time evolution of Eq.(\ref{circular}) in the form of an animation can be found in the Appendix.

\section{Rotating waves along a confocal AJTJ}

The analysis of a confocal AJTJ (drawn in the bottom panel of Fig.~\ref{drawings}) is better carried out using the (planar) elliptic coordinates system $(\nu,\tau)$, where $\nu\geq0$ and $\tau\in[-\pi,\pi]$ are the radial and angular variables, respectively. Once the position of the foci $(\pm c,0)$ is given, all possible confocal ellipses are uniquely identified by a characteristic value, $\nu_c$. In the limit of a vanishing eccentricity, the foci of the ellipse collapse to a point at the origin (i.e., $c\to0$) and the ellipse turns into a circle. At the same time, $\cosh\nu_c$ diverges, while the product $c \cosh\nu_c$ remains finite and tends to the radius, $r$, of the circle.

\medskip

\noindent If the confocal AJTJ is delimited by two closely spaced ellipses identified by the characteristic values $\nu_{in}$ and $\nu_{out}\gtrapprox \nu_{in}$, they have almost the same aspect ratio, $\rho$ (defined as the ratio of the length of the minor axis to the length of the major axis), and so the same eccentricity $e^2\equiv 1- \rho^2<1$. When the common focal length, $c$, of the ellipses tends to zero, then the annulus eccentricity vanishes and the variable-width confocal AJTJ reduces to the constant-width circular AJTJ. For $\Delta\nu\equiv \nu_{out}-\nu_{in}<<1$, the width of the confocal annulus varies as \cite{JLTP16b}:

\begin{equation}
\Delta w(\tau)=c\mathcal{Q}(\tau)\,\Delta\nu,
\label{width}
\end{equation}

\noindent where $\mathcal{Q}(\tau)$ is the elliptic $\pi$-periodic scale factor, defined as $\mathcal{Q}(\tau) \equiv \sqrt{(\cosh2\bar{\nu} + \cos2\tau)/2}$, that oscillates between $\sinh\bar{\nu}$ and $\cosh\bar{\nu}$ with $\bar{\nu}\equiv (\nu_{in}+\nu_{out})/2$. The parameter $\bar{\nu}$ is strictly related to the aspect ratio, $\rho=\tanh \bar{\nu}\leq1$, as well as to the annulus eccentricity \cite{JPCM16}, as $e^2=\sech^2 \bar{\nu}\leq1$. Therefore, large eccentricities correspond to small $\bar{\nu}$ values. A narrow confocal annulus is uniquely identified by the geometrical parameters $c$ and $\bar{\nu}$.

\medskip

\noindent In the small width approximation, $\Delta w_{max}=c \Delta \nu  \cosh \bar{\nu} << \lambda_J$, the \Jos phase of a confocal AJTJ does not depend on $\nu$ and the system is one-dimensional. It has been derived that the radially independent \Jos phase, $\phi(\tau,\hat{t})$, of a confocal AJTJ in the presence of a spatially homogeneous in-plane magnetic field ${\bf H}$ of arbitrary orientation, $\bar{\theta}$, relative to the $y$-axis, obeys a modified and perturbed sine-Gordon equation with a space dependent effective Josephson penetration length inversely proportional to the local junction width \cite{JLTP16b}:

\begin{equation}
\left[\frac{\lambda_J}{c\,\mathcal{Q}(\tau)}\right]^2 \phi_{\tau\tau} - \phi_{\hat{t}\hat{t}}-\sin \phi =\alpha \phi_{\hat{t}} - \gamma + F_h(\tau),
\label{psge}
\end{equation}

\noindent where $\gamma$ is the normalized density of the constant bias current and 

\begin{equation}
F_h(\tau)\equiv h\Delta \frac{\cos\bar{\theta}\cosh\bar{\nu} \sin\tau-\sin\bar{\theta}\sinh\bar{\nu}\cos\tau }{\mathcal{Q}^2(\tau)}
\label{Fh}
\end{equation}

\noindent is an additional forcing term proportional to the applied magnetic field; here, $h\equiv H/J_c c$ is the normalized field strength for treating long CAJTJs and $\Delta$ is a geometrical factor which sometimes has been referred to as the coupling between the external field and the flux density of the annular junction \cite{Gronbech}. As usual, the $\alpha$ term in Eq.(\ref{psge}) accounts for the quasi-particle shunt loss. Eq.(\ref{psge}) can be classified as a perturbed and modified sine-Gordon equation in which the perturbations are given by the dissipation and driving fields, while the modification is represented by an effective local $\pi$-periodic \Jos penetration length, $\Lambda_J(\tau)\equiv \lambda_J/Q(\tau)= c \lambda_J \Delta \nu /\Delta\!W(\tau)$, inversely proportional to the annulus width. It is worth to point out that this $\Lambda_J$ variation stems from the variable junction width and cannot be modeled in terms of a spatially varying $\lambda_J$ in uniform-width JTL treated in Refs.(\cite{Sakai},\cite{Petras}); nevertheless, in the time independent case, it happens to be equivalent to a change in the $J_c$ of a uniform-width JTL \cite{Semerdzhieva}. Notably, the PDE of a confocal AJTJ does not differ from that of a circular one in Eq.(\ref{PDEring}) provided that \cite{note2} the product $c\mathcal{Q}$ is replaced by the mean radius, $r$, of the ring and the tangential elliptic coordinate $\tau$ is changed into the polar angle $\theta$. Furthermore, in the limits $c\to0$ and $\nu\to\infty$, the elliptic coordinates $(\nu,\tau)$ reduce to polar coordinates $(r,\theta)$ and magnetic force in Eq.(\ref{Fh}) reduces to the sinusoidal forcing term of Eq.(\ref{PDEring}). At last, to take into account the possible vortices (or fluxons) trapped in the confocal AJTJ due to flux quantization in one of the superconducting electrodes, Eq.(\ref{psge}) is supplemented by the periodic boundary conditions in Eqs.(\ref{periodic1}) and (\ref{periodic2}) with $\theta$ replaced by $\tau$.


\subsection{Small amplitude oscillations in a confocal AJTJ}

\noindent No analytical solutions exist to Eq.(\ref{psge}) even with the right-hand side set to zero. However, extensive numerical simulations carried out on long confocal AJTJs having an aspect ratio $\rho=1/2$ and $1/4$ showed that dynamic solutions to the PDE exist in the form of rotating $2\pi$-kinks (fluxons) and/or circulating waves \cite{JLTP18,SUST18}. In this section we will investigate the analytical solutions to the autonomous PDE which have the form of small-amplitude oscillations. With this approximation, and $\sin\phi\approx\phi$, the left-hand side of Eq.(\ref{psge}) becomes:

\begin{equation}
\left[\frac{1}{\hat{c}\,\mathcal{Q}(\tau)}\right]^2 \phi_{\tau\tau} - \phi_{\hat{t}\hat{t}}- \phi=0,
\label{psge5}
\end{equation}

\noindent where we have introduced the normalized focal distance $\hat{c}\equiv c/\lambda_J$. This linear PDE can be solved by variable separation assuming that $\phi(\tau,\hat{t})=  \Phi(\tau) \exp[-\hat{\imath} (\hat{\omega} \hat{t}-\delta)]$. Inserting $\phi$ into Eq.(\ref{psge5}) and recalling that $\mathcal{Q}^2(\tau)=(\cosh2\nu + \cos2\tau)/2$, after some algebraic manipulations, the equation can be cast in the form of an angular Mathieu differential equation for $\Phi(\tau)$:

\begin{equation}
 \frac{d^2\Phi}{d\tau^2} + \left(a-2q\cos2\tau\right) \Phi(\tau)=0,
\label{mathieu}
\end{equation}

\noindent where $a=\hat{c}^2 (\hat{\omega}^2-1)  \cosh2\bar{\nu}/2\geq0$ and $q=-\hat{c}^2 (\hat{\omega}^2-1) /4\leq0$. With $q=0$, Eq.(\ref{mathieu}) admits periodic solutions only if $\sqrt{a}$ is an integer. At variance with the Mathieu equation in Eq.(\ref{mathieu0}) for a circular AJTJ in the presence of a uniform magnetic fields, now the coefficients $a$ and $q$ both depend on the frequency $\omega$. Therefore, once the geometrical parameters $c$ and $\bar{\nu}$ are given, the eigenfrequencies $\hat{\omega}_{a,m}$ and $\hat{\omega}_{b,m}$ for the even and odd parts of the solutions, $\Phi_m(\tau)$, with periodicity $2\pi/m$ have to be determined by solving, respectively, the equations \textsl{MathieuCharacteristicA}$[m, -\hat{c}^2 (\hat{\omega}_{a,m}^2-1) /4]=\hat{c}^2 (\hat{\omega}_{a,m}^2-1)  \cosh2\bar{\nu}/2$ and \textsl{MathieuCharacteristicB}$[m, -\hat{c}^2 (\hat{\omega}_{b,m}^2-1) /4]=\hat{c}^2 (\hat{\omega}_{b,m}^2-1)  \cosh2\bar{\nu}/2$. Then, following the line of reasoning in Section~\ref{sec:sexion}, the general solution to Eq.(\ref{psge5}) having a $\tau$-periodicity equal to $2\pi/m$ is:


\begin{figure}[tb]
\centering
\includegraphics[height=6cm,width=10cm]{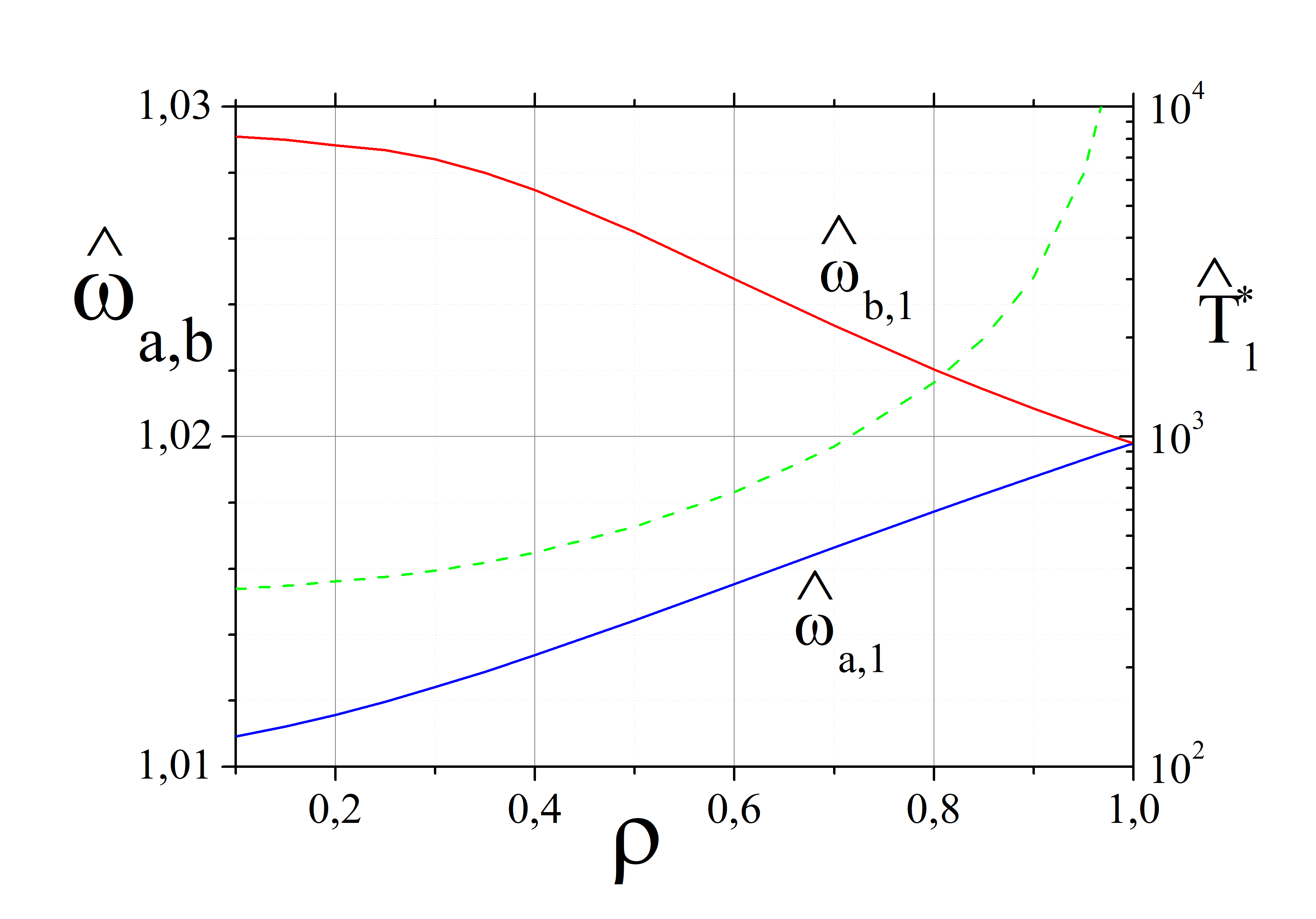}
\caption{(Color online) The dependence of the first-mode eigenfrequencies, $\hat{\omega}_{a,1}$ and $\hat{\omega}_{b,1}$, on the aspect ratio, $\rho$, for a confocal AJTJ having normalized circumference $\ell=10\pi$; as $\rho\to1$, the frequencies degenerate to the value $1.0198$, as expected for a circular AJTJ with the same circumference in the absence of a magnetic field (see Fig.~\ref{omegaab} for $h=0$). The dashed line represents the wave periodicity, $\hat{T}_1^*(\rho)\equiv 2\pi /(\hat{\omega}_{b,1}-\hat{\omega}_{a,1})$, and refers to the right (logarithmic) scale.}
\label{omegaabConf}
\end{figure}


\begin{equation}
\tilde{\phi}_{m,\bar{\nu}}(\tau,\hat{t})=A_m C(a_m,q_m,\tau) \sin\hat{\omega}_{a,m}\hat{t}+B_m S(b_m,p_m,\tau) \cos\hat{\omega}_{b,m}\hat{t},
\label{phimnu}
\end{equation}

\noindent where $|A_m| \cong |B_m|<<1$, $a_m=\hat{c}^2 (\hat{\omega}_{a,m}^2-1)  \cosh2\bar{\nu}/2$, $b_m=\hat{c}^2 (\hat{\omega}_{b,m}^2-1)  \cosh2\bar{\nu}/2$, $q_m=-\hat{c}^2 (\hat{\omega}_{a,m}^2-1)/4$ and $p_m=-\hat{c}^2 (\hat{\omega}_{q,m}^2-1) /4$. The solution in Eq.(\ref{phimnu}) has been obtained by imposing $Re[\tilde{\phi}_{m,\bar{\nu}}(0,0)]=0$ and, as previously commented, assuming that it is a small amplitude oscillation that does not maintain its shape and periodically changes its direction of propagation. As the confocal annulus tends to a circular annulus of radius $r$, i.e., in the limits $c\to0$ and $c \cosh \nu=c \sinh \nu \to r$, then,  $a_m=b_m=\hat{r}^2 (\hat{\omega}_m^2-1)=m^2$, $p_m=q_m=0$ and $\hat{\omega}_{a,m} =\hat{\omega}_{b,m}=\hat{\omega}_{m}= \sqrt{1+(m/\hat{r})^2}$; therefore, as expected, the solution in Eq.(\ref{phimnu}), with $B_m=-A_m$, reduces to a circulating plane wave,
$\tilde{\phi}_{m,\infty}(\tau,\hat{t})=A_m \sin (m\tau -\hat{\omega}_{m} \hat{t})$, already found in Eq.(\ref{circularsmall}).
\medskip
\noindent The mean perimeter of a confocal elliptic annulus is $L= 4c \cosh\bar{\nu}\,\texttt{E}[\sech^2 \bar{\nu}]$. Then the normalized perimeter or electric length, $\ell=L/\lambda_J$, of the CAJTJ of a given aspect ratio grows linearly with the distance, $c$, of a focus from the center. The solid lines in Fig.~\ref{omegaabConf} show the (weak) dependence of the first-mode eigenfrequencies $\hat{\omega}_{a,1}$ and $\hat{\omega}_{b,1}\geq \hat{\omega}_{a,1}$ on the aspect ratio, $\rho$, for a confocal AJTJ having a fixed circumference $\ell=10\pi$; as $\rho\to1$, the frequencies degenerate to the value $1.0198$, as expected for a circular AJTJ with the same circumference in the absence of a magnetic field (see Fig.~\ref{omegaab} for $h=0$). The dashed line represents the wave periodicity, $\hat{T}_1^*(\rho)\equiv 2\pi /(\hat{\omega}_{b,1}-\hat{\omega}_{a,1})$, and refers to the right (logarithmic) scale.


\section{Conclusions}

We have investigated the propagation of EM waves in annular \Jos tunnel junctions having different geometrical configurations, namely, the well-known circular geometry and the more general confocal elliptic geometry. In both cases, perturbed sine-Gordon equations govern the junction phase dynamics whose spatial dependencies, in the limit of small amplitude oscillations, are found to be periodic solution to the Mathieu equations. Qualitatively similar dispersion relations have been predicted in circular JTJs when a magnetic field is applied in the plane of the junction and in confocal AJTJs with no applied field. More specifically, for each discrete mode $m$, that corresponds to a wavelength equal to the annulus circumference divided by $m$, two eigenfrequencies exist that are related to the even and odd spatial dependence of the wave. As a result of this frequency split, the traveling wave is given by the superposition of two standing waves with the same wavelengths but different oscillation periods. Therefore, the wave profile and the velocity of the wave front are not permanent, but undergo periodic changes. Although the main manifestation of EM waves is represented by their nonlinear interaction with traveling solitary magnetic flux quanta in long JTJs our analysis applies to junction of any physical and electric length. All our finding on AJTJs have been successfully double-checked by detailed numerical simulations.



\renewcommand{\theequation}{A-\arabic{equation}}
\setcounter{equation}{0}  
\setcounter{subsection}{0}  
\section*{Appendix - \textit{Mathematica} code for the animation of Eq.(21)}

\noindent The purpose of this appendix is to illustrate a \textit{Mathematica} (www.wolfram.com) notebook that displays the time evolution of the Josephson phase for an unbiased and lossless circular \ann \Jos tunnel \jun in the presence of an in-plane magnetic field. The code calculates $\tilde{\phi}_{m,h}$ as given in Eq.(\ref{phimh}) that corresponds to the $m$-mode of a rotating EM wave. The notation of the code are those in the text. However, to make the code fully copyable as a plain text and pastable as a \textit{Mathematica} notebook, the accents have been eliminated, the subscripts have been raised and the Greek symbols have been replaced by their names; as an example, ``$\tilde{\phi}_{m,h}$'' become ``phimh''.

\begin{footnotesize}
\begin{verbatim}


(*                   first line of the notebook                       *)

m=2;             (* integer mode number *)

r=5.;            (* normalized radius *)

re=r/Sqrt[1+r^2];           (* effective radius *)

q[h_]:=(h^2 re^2 r^2)/8     (* q parameter *)

omegaam[h_]:=Sqrt[MathieuCharacteristicA[m,(h^2 re^2 r^2)/8]/ r^2+1-h^2 re^2/4]

omegabm[h_]:=Sqrt[MathieuCharacteristicB[m,(h^2 re^2 r^2)/8]/ r^2+1-h^2 re^2/4]

Plot[{omegaam[h],omegabm[h]}, {h,0,1},
PlotStyle->{Blue,Red},AxesOrigin->{0,0.8},
PlotLabel->Style["frequencies vs magnetic field",FontSize->20]];
(* frequencies plot *)

Plot[{2Pi/(omegaam[h]-omegabm[h])}, {h,0.1,0.3},
PlotStyle->{Blue},AxesOrigin->{0,0.8},
PlotLabel->Style["periodicity vs magnetic field",FontSize->20]];
(* periodicity plot *)

(* usefull evaluations for a specific value, hh, of the 
normalized magnetic field *)

hh=0.9;    (* normalized magnetic field value *)

qq=q[hh];                  

am=MathieuCharacteristicA[m,qq[h]]/.{h->hh};

bm=MathieuCharacteristicB[m,qq[h]]/.{h->hh};

omegaam[hh];

omegabm[hh];

T=2Pi/(omegaam[hh]-omegabm[hh]) ;  (* wave periodicity *)

(*		expression of the Josephson phase in Eq.21	 *)
phimh[A_,B_,h_,theta_,t_]:=
A MathieuC[MathieuCharacteristicA[m,(h^2 re^2 r^2)/8],(h^2 re^2 r^2)/8,theta]Sin[omegaam[h]t]+
B MathieuS[MathieuCharacteristicB[m,(h^2 re^2 r^2)/8],(h^2 re^2 r^2)/8,theta]Cos[omegabm[h]t]

(*	    Animation command			 *)
Animate[Plot[phimh[A,B,h,theta,t]/.{A->0.2,B->-0.2,h->hh},{theta,-Pi,Pi},
PlotRange->0.5,Frame->True,PlotStyle->{Red},PlotLabel->Style["phimh(theta)",FontSize->20],
GridLines->Automatic,ImageSize->Scaled[0.5]],{t,0,300},AnimationRunning->False,AnimationRate->1]

(*                   last line of the notebook                       *)

    \end{verbatim}
		\end{footnotesize}
		
\begin{figure}[tb]
\centering
\includegraphics[height=5cm,width=9cm]{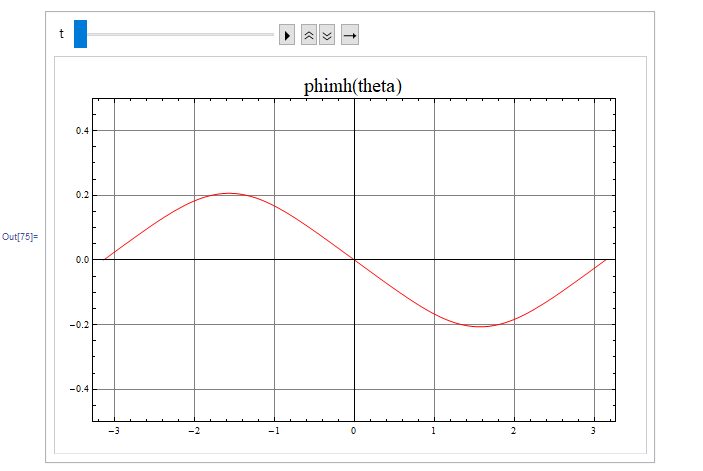}
\caption{(Color online) Animation screenshot.}
\label{animation}
\end{figure}

\noindent Once the \textit{Mathematica} dynamics has been enabled the code provides a frame with the phase profile at $t=0$, $\tilde{\phi}_{m,h}(\theta,\!\hat{t})$ with $\theta$ in the interval $[-\pi,\pi]$, as shown in Figure~\ref{animation}. By clicking on the Start/Stop button, the animation begins and lasts for 600 units of time, as for the given settings, the wave periodicity,$T$, of the first mode is equal to about 587 time units; then the wave will invert his velocity at $t\approx300$. As the operating magnetic field increases above the set value $hh=03$, the wave becomes more and more non-linear. For the second mode, $m=2$, one should set the operating magnetic field value around $hh=0.9$ or above. Special buttons are available to accelerate, slow down or invert the animation speed.

\section*{Acknowledgments}
\noindent The author wishes to thank J. Mygind for stimulating discussions and for a critical reading of the manuscript. 
\newpage


\begin{thebibliography}{00}
\bibitem{notePsih} \begin{footnotesize} With $\psi_h(\theta) = 2\arcsin\,\left[(h_e/2) \sin(\bar{\theta}-\theta)\right]$ and assuming $\bar{\theta}=0$, it is: $\sin \psi_h(\theta)=2\sin\left\{\arcsin\,\left[(h_e/2) \sin(-\theta)\right]  \right\}\cos\left\{\arcsin\,\left[(h_e/2) \sin(-\theta)\right]  \right\}=-h_e \sin \theta \sqrt{1-(h_e/2)^2 sin^2 \theta} \approx - h_e \sin \theta$, $\cos \psi_h(\theta)=1-2\sin^2 \left\{\arcsin\,\left[(h_e/2) \sin(-\theta)\right]  \right\}=1-2\left[(h_e/2) \sin\theta\right]^2 =1-(h_e/2)^2+(h_e/2)^2 \cos 2 \theta$, $d\psi_h(\theta)/d\theta=-h_e \cos \theta/\sqrt{1-(h_e/2)^2 sin^2 \theta} \approx -h_e \cos \theta$ and $d^2\psi_h(\theta)/d\theta^2=h_e [1-(h_e/2)^2]\sin \theta/[1-(h_e/2)^2 sin^2 \theta]^{3/2} \approx h_e \sin \theta$.\end{footnotesize}
\bibitem{note2} \begin{footnotesize} In the limit of vanishing eccentricity ($e^2\simeq0$ and $\cosh2\bar{\nu}>>1$), it is $c^2 \mathcal{Q}^2(\tau)= c^2 \left(\cosh2\bar{\nu}+\cos2\tau\right)/2 \approx c^2\cosh2\bar{\nu}/2=c^2 (\cosh^2\bar{\nu}+\sinh^2\bar{\nu})/2\simeq
c^2 \cosh^2\bar{\nu}\approx r^2$.\end{footnotesize}
\end{thebibliography}
\end{document}